\documentclass[
twocolumn,
superscriptaddress,
citeautoscript
 amsmath,amssymb,
 aps,
]{revtex4-2}

\usepackage{graphicx}
\usepackage{dcolumn}
\usepackage{bm}
\usepackage{amsmath}
\usepackage{amsthm}
\usepackage{amsfonts}
\usepackage{amssymb}
\usepackage{mathtools}
\usepackage[utf8]{inputenc}
\usepackage{blkarray}
\usepackage{pgfplots}
\usepackage{tikz}
\usetikzlibrary{shapes.geometric, automata, positioning, arrows,patterns}
\usepackage{float}
\pgfplotsset{compat=1.16}
\newtheorem{theorem}{Theorem}[section]

\begin{document}

\tikzstyle{block} = [draw, fill=white, rectangle, 
    minimum height=3em, minimum width=4.5em]
\tikzstyle{input} = [coordinate]
\tikzstyle{output} = [coordinate]

\preprint{APS/123-QED}

\title{The channel capacity of the ribosome}

\author{Daniel A. Inafuku}
 \email{inafuku2@illinois.edu}
 \affiliation{Department of Physics, University of Illinois Urbana-Champaign, Urbana, IL 61801}
 \author{Kay L. Kirkpatrick}
 \email{kkirkpat@illinois.edu}
 \affiliation{Department of Physics, University of Illinois Urbana-Champaign, Urbana, IL 61801}
 \affiliation{Department of Mathematics, University of Illinois Urbana-Champaign, Urbana, IL 61801}
 \author{Onyema Osuagwu}
 \email{onyema.osuagwu@morgan.edu}
 \affiliation{Electrical and Computer Engineering Department, Morgan State University, Baltimore, MD 21251}
\affiliation{Cybersecurity Assurance and Policy Center, Morgan State University, Baltimore, MD 21251}
\author{Qier An}
 \email{aan@ucsb.edu}
 \affiliation{Department of Physics, University of California, Santa Barbara, Santa Barbara, CA 93106}
\author{David A. Brewster}
 \email{dbrewster@g.harvard.edu}
  \affiliation{Department of Mathematics, University of Illinois Urbana-Champaign, Urbana, IL 61801}
 \affiliation{John A. Paulson School of Engineering and Applied Sciences, Harvard University, Cambridge, MA 02138}
\author{Mayisha Zeb Nakib}
 \email{mnakib2@illinois.edu}
 \affiliation{Department of Physics, University of Illinois Urbana-Champaign, Urbana, IL 61801}

\begin{abstract}
Translation is one of the most fundamental processes in the biological cell. Because of the central role that translation plays across all domains of life, the enzyme that carries out this process, the ribosome, is required to process information with high accuracy. This accuracy often approaches values near unity experimentally. In this paper, we model the ribosome as an information channel and demonstrate mathematically that this biological machine has information-processing capabilities that have not been recognized previously. In particular, we calculate bounds on the ribosome's theoretical Shannon capacity and numerically approximate this capacity. Finally, by incorporating estimates on the ribosome's operation time, we show that the ribosome operates at speeds safely below its capacity, allowing the ribosome to process information with an arbitrary degree of error. Our results show that the ribosome achieves a high accuracy in line with purely information-theoretic means.

\end{abstract}

\maketitle

\section{\label{sec:level1}Introduction}

The ribosome is a Brownian nano-machine that assembles proteins from codon sequences in messenger RNA (mRNA, and codons are nucleotide triplets), matching each codon to an anticodon and through that to an amino acid by a kind of look-up table (i.e., the genetic code) in the physical form of transfer RNA (tRNA) \cite{dashti14, prabhakar17}. After joining the codon with its anticodon tRNA, the ribosome catalyzes the peptide bond formation, producing an amino acid string that folds into a protein.

A ribosome is a one-way, almost deterministic, finite transducer (in the terminology of Aho \cite{aho69}): almost deterministic in the sense that rare errors occur approximately once in 1,000 to 10,000 codons \cite{edelmann77, drummond09, parker89, ogle05, kramer07, alberts22}. Ribosomes process between about 99.9$\%$ to 99.99$\%$ of codons accurately, thanks to proof-reading mechanisms, and errors often result in premature abandonment of translation. Ribosomes usually halt correctly at stop codons but occasionally get stalled if a stop codon is missing, damaged, or misread. Such stalling can be deadly for a cell, but there are mechanisms in eukaryotes for rescue \cite{ishimura14, james16, zeng17}.

In addition, the ribosome is a memoryless finite-state machine having 64 codon symbols and 20 amino acid states: memoryless because the ribosome's current state determines its next action, and that next action is energetically favorable \cite{savir13}. Because it links the amino acids in an ordered chain, there are combinatorially many possible output proteins. In theory, a ribosome could make more than $20^{50}$ outputs (50-2,000 amino acids being the length of a typical protein \cite{alberts22}, and some can potentially reach 38,000 amino acids long \cite{bang01, kruger11}); although in practice, the ribosome is limited by the information it is fed by the mRNA sequences. 

Ribosomes operate quickly, translating a codon in about 50 milliseconds and producing a typical-length polypeptide on the order of minutes \cite{prabhakar17, bremer08, alberts22}. The polypeptide then folds into its functional protein form, with the fastest folding times being on the order of microseconds \cite{kubelka04}. 

The natural interpretation of protein synthesis as a process of information transmission is widespread and may contribute to our understanding of the ribosome's simultaneously high accuracy and speed. Applications of information theory are numerous: efforts have been made at the neuron, network \cite{juarrero99}, and system levels in a variety of ways with names such as information bottleneck \cite{tishby99}, information distortion \cite{dimitrov03}, effective information \cite{hoel17}, consistent information \cite{corominasmutra14}, teleosemantic information \cite{cao11}, and positional information \cite{tkacik15, dubuis13, petkova19, sokolowski22}. But there is no consensus yet on which are the most useful interpretations, and they are all problematic \cite{keller09}.

Calculations of information-theoretic quantities focusing generally on gene expression and protein synthesis have been conducted previously using specially constructed channel matrices \cite{yockey74, yockey05, djordjevic12}. We build on these results by introducing a novel channel matrix for the ribosome and show that it operates at rates below its channel capacity and satisfies the hypotheses of Shannon's Noisy Channel Coding theorem, allowing the ribosome to transmit information with an arbitrary degree of error. We do so by modeling the ribosome as an information channel, calculating bounds on the ribosome's channel capacity, and comparing the capacity with experimentally determined translation rates. These results provide explanations for the ribosome's high accuracy despite its high translation rate.

\section{\label{sec:level1}The ribosome works within its channel capacity}

We view the ribosome as a discrete memoryless channel: the input message $X^*$ is mRNA, and the output $Y^*$ is the resulting polypeptide. As mentioned above, the ribosome has an accuracy of about 99.9$\%$ to 99.99$\%$. Why is this high level of accuracy possible? Shannon's Noisy Channel Coding theorem sets the channel capacity $\mathcal{C}$ as the maximum rate at which a channel can transmit information with arbitrarily low error \cite{cover06}. The capacity is defined as
\begin{equation}
    \mathcal{C} := \sup_{p_X}I(X;Y),
    \label{eq:1}
\end{equation}

\noindent where $I(X;Y)$ is the mutual information of random variables $X$ (input) and $Y$ (output), and the supremum is taken over all input distributions $p_X$. In Section II.A, we give explicit bounds on Eq.~(\ref{eq:1}). For the remainder of this paper, ``$\log$" denotes the base-2 logarithm so that the units of information are bits, unless stated otherwise.

While we acknowledge that several ribosome variants exist---for example, there are structural differences between the ribosomes of prokaryotes and eukaryotes---we also recognize that the fundamental informational function of translation across these different variants remains the same \cite{clark19}. Therefore, we focus on the ribosome's basic translational process, the conversion of information contained in mRNA's codon string into a polypeptide string (Fig. 1).

\subsection{\label{sec:level2} Bounding the Ribosome's Capacity}

\begin{figure}
    \centering
    \fboxsep=4mm
\fbox{
\begin{tikzpicture}[->, node distance=3cm,>=latex']
\tikzset{->, >=stealth',node distance=3cm, every state/.style={thin, fill=gray!10},initial text=$ $,}

    \node[block] (D) {DNA};
    \node[block, right of=D, node distance=5cm] (M) {mRNA $X^*$};
    \node[block, below of=M, node distance=2cm] (R) {ribosome $p(y|x)$};
    \node [block, below of=R, node distance=2cm] (P) {polypeptide $Y^*$};
    \node [block, left of=P, node distance=5cm] (F) {protein $\Tilde{Y}$};
    
   \draw 
         (D) edge[above] node{transcription} (M)
         (M) edge[right] node{} (R)
         (R) edge[right] node{} (P)
         (P) edge[above] node{folding} (F);

\end{tikzpicture}
}
    \caption{The ribosome as an information-theoretic channel: input string $X^*$ is translated into an output string $Y^*$, which folds into protein $\widetilde{Y}$.}
    \label{fig:my_label}
\end{figure}
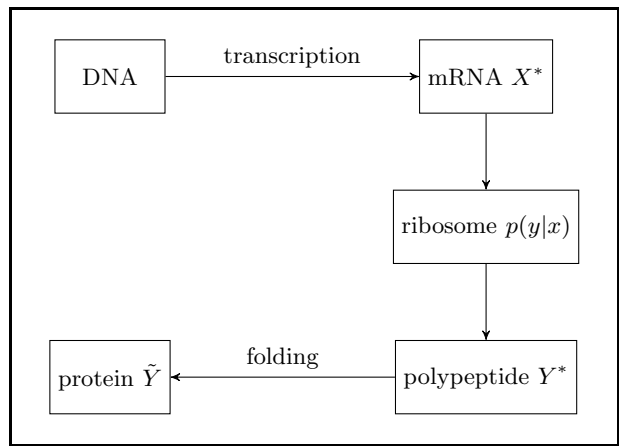

We first note that there are $4^3=64$ codons, each requiring $\log64=6$ bits to be specified as triplets from the input alphabet $\mathcal{X}:= \{ \text{A}, \text{C}, \text{G}, \text{U}\}^3$. Additionally, the output set is the alphabet $\mathcal{Y}:=  \{\text{Met}, \, \text{Leu}, \, \ldots, \, \text{Ser}, \,\text{Stop}\}$, which includes all 20 standard proteinogenic amino acids and the ``Stop" symbol. Therefore, each amino acid is specified by $\log21 \approx 4.3923$ bits.

We model the ribosome directly as an information channel by specifying its conditional probability distribution $p(y|x)$:
\begin{equation}
    p(y|x)= 
    \begin{cases}
    1-r \enskip , &  y=\mathcal{G}(x)\\
    \, \, \, \, \frac{r}{20} \enskip , &  y \neq \mathcal{G}(x),\\
    \end{cases}
    \label{eq:2}
\end{equation}

\noindent where $\mathcal{G}: \mathcal{X} \to \mathcal{Y}$ is the standard genetic code, and $r \in [0,1]$ is the probability of error. $\mathcal{G}$ is constructed from a table of the standard genetic code \cite{clark19}. For example, $\mathcal{G}(\text{AUG})=\text{Met}$, where Met is the amino acid methionine. A diagram of the ribosome as an information channel is given by Fig. 2.

According to Eq.~(\ref{eq:2}), the ribosome correctly matches a codon $x \in \mathcal{X}$ to its corresponding amino acid $y \in \mathcal{Y}$ according to $\mathcal{G}$ (i.e., $y=\mathcal{G}(x)$) with some (preferably large) probability $1-r$. Conversely, there is a (preferably small) probability $r$ that the ribosome performs an incorrect match (i.e., $y \neq \mathcal{G}(x)$). In this case, our model assumes that the error probability $r$ is distributed equally over all 20 possible incorrect outputs in $\mathcal{Y}$, i.e., for a fixed codon $x$, each $y \neq \mathcal{G}(x)$ has a probability $\frac{r}{20}$. These conditions ensure the proper normalization of $p(y|x)$.

Eq.~(\ref{eq:2}) resembles the well-studied $q$-ary symmetric channel \cite{shokrollahi04}, except that here, the input and output alphabets are different (codons vs. amino acids), whereas the input and output alphabets of the $q$-ary symmetric channel are identical to each other. Moreover, our channel defined by Eq.~(\ref{eq:2}) depends on an external function, namely, the genetic code $\mathcal{G}$. These new features necessitate a new calculation in our ribosomal context.

Eq.~(\ref{eq:2}) is a $64 \times 21$ channel matrix and contains a sufficiently large degree of asymmetry such that a closed form of the capacity $\mathcal{C}$ is difficult to obtain. However, one can still bound $\mathcal{C}$, and we do so here. As a quick estimate, we immediately see that the channel capacity is bounded above:
\begin{equation}
0 \leq \mathcal{C} \leq \min \{\log|\mathcal{X}|, \log|\mathcal{Y}|\}=\log{21} \approx 4.3923,
\label{eq:3}
\end{equation}

\noindent where the upper bound is obtained by maximizing the entropy of $X$ and $Y$, respectively \cite{cover06}.

\begin{figure}[t]
    \begin{center}
    \begin{tikzpicture}
    \tikzset{->, >=stealth',node distance=3cm, every state/.style={thin, fill=gray!10},initial text=$ $,}

    \node[state] (S1) {UUU};
    \node[state, below of=S1, yshift=1cm] (S2) {UUC};
    \node[state, below of=S2, yshift=1cm] (S3) {UUA};
    \node[text height=0.5cm] at (0.02,-4.9) {$\mathbf{\vdots}$};
    \node[text height=0.5cm] at (0.02,-5.7) {$\mathbf{\vdots}$};
    \node[state, below of=S3, yshift=0cm] (S4) {GGC};
    \node[state, below of=S4, yshift=1cm] (S5) {GGA};
    \node[state, below of=S5, yshift=1cm] (S6) {GGG};
    
    \node[state, right of=S1, , xshift =3.5cm, yshift=-1.5cm] (T1) {Phe};
    \node[state, right of=S1, xshift=3.5cm, yshift=-3.5cm] (T2) {Leu};
    \node[text height=0.5cm] at (6.5,-4.9) {$\mathbf{\vdots}$};
    \node[text height=0.5cm] at (6.5,-5.7) {$\mathbf{\vdots}$};
    \node[state, right of=S1, xshift=3.5cm, yshift=-7.5cm] (T3) {Arg};
    \node[state, right of=S1, xshift=3.5cm, yshift=-9.5cm] (T4) {Gly};
    
    \draw   
        (S1) edge[above] node{} (T1)
        (S1) edge[above right, dashed,->] node{} (T2)
        (S1) edge[above right, dashed,->] node{} (T3)
       (S1) edge[below left, dashed,->] node{} (T4)
       
       (S2) edge (T1)
       
       (S3) edge (T2)
       
       (S4) edge (T4)
       
       (S5) edge (T4)
       
       (S6) edge (T4);

    \end{tikzpicture}
    \caption{Transmission diagram for the ribosome. There are 64 input symbols (codons, left) and 21 output symbols (amino acids plus the ``Stop" symbol, right), although most symbols are hidden for clarity. Solid arrows indicate ``correct" transmissions (i.e., $y=\mathcal{G}(x)$), and dashed arrows represent ``incorrect" transmissions (i.e., $y\neq \mathcal{G}(x)$). Only UUU's incorrect transmissions are depicted here. Synonymous codons are mapped to the same amino acid---e.g., codons GGC, GGA, and GGG are all mapped by $\mathcal{G}$ to the amino acid glycine (Gly). Synonymous codons represent the degeneracy of $\mathcal{G}$, which contributes to the ribosomal channel's asymmetry.}
    \label{fig:my_label}
    \end{center}
\end{figure}
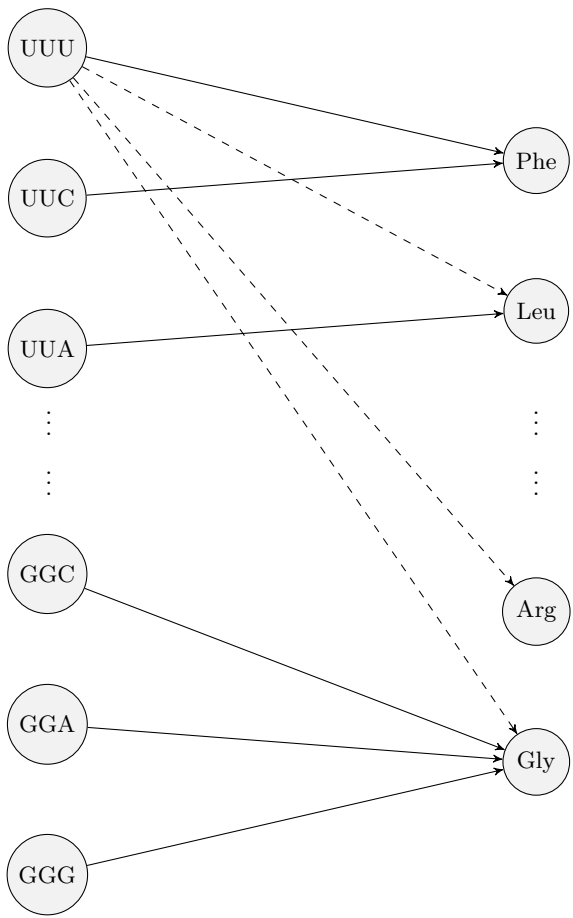

We improve the capacity's lower bound by an explicit calculation using Eq.~(\ref{eq:1}). Combining this calculation with Eq.~(\ref{eq:3}), we obtain the following theorem, whose proof we outline in the appendix.

\begin{theorem}
The channel capacity $\mathcal{C}$ of the ribosome in bits/use satisfies 
$g(r) \leq \mathcal{C} \leq \log21$, where $g(r)$ is given by
\begin{align}
    g(r)  &:= \frac{1}{64} \bigg\{2q\log \frac{1280q}{43r+20}+\frac{63r}{10} \log \frac{64r}{43r+20} \nonumber \\
         & + 18q\log \frac{640q}{11r+20}+\frac{279r}{10} \log \frac{32r}{11r+20} \nonumber \\
         & + 6q\log \frac{1280q}{r+60}+\frac{61r}{10} \log \frac{64r}{r+60}\nonumber \\
         & + 20q\log \frac{64q}{4-r}+15r\log \frac{16r}{5(4-r)}\nonumber \\
         & + 18q\log \frac{640q}{60-31r}+\frac{87r}{10} \log \frac{32r}{60-31r} \bigg\}
\end{align}
\noindent and $q:=1-r$.
\label{eq:4}
\end{theorem}

\begin{figure}[t]
    \centering
    \begin{tikzpicture}
\begin{semilogxaxis}[domain=10^-6:1, samples=200, xlabel = $r$,ylabel={Information (bits/use)},xmajorgrids=true,ymajorgrids=true, every axis plot/.append style={very thick}, legend style={at={(.1,.2)}, anchor=south west}]
\addplot[color=black]{(2*(1-x)*log2((1280*(1-x))/(43*x+20)) + (63/10)*x*log2(64*x/(43*x+20))+18*(1-x)*log2((640*(1-x))/(11*x+20)) + (279/10)*x*log2(32*x/(11*x+20))+6*(1-x)*log2((1280*(1-x))/(x+60)) + (61/10)*x*log2(64*x/(x+60))+20*(1-x)*log2((64*(1-x))/(4-x)) + 15*x*log2(16*x/(5*(4-x)))+18*(1-x)*log2((640*(1-x))/(60-31*x)) + (87/10)*x*log2(32*x/(60-31*x)))/64};
\addplot[dashed][domain=10^-6:0.183]{log2(61)-1.68+ 6.509*x*log2(0.4594*x)};
\addplot[dotted]{log2(21)};
\legend{$g(r)$,$I(r)$,$\log{21}$}
\end{semilogxaxis}
\end{tikzpicture}
    \caption{A linear-log plot showing the capacity's lower bound $g(r)$ (solid), Djordjevic's maximization $I(r)$ of Yockey's mutual information (dashed), and the capacity's upper bound in Theorem II.1 (dotted) as functions of error probability $r$. The space between the solid and dotted curves represents the region where we predict that the capacity may lie.}
   \label{fig:my_label}
\end{figure}
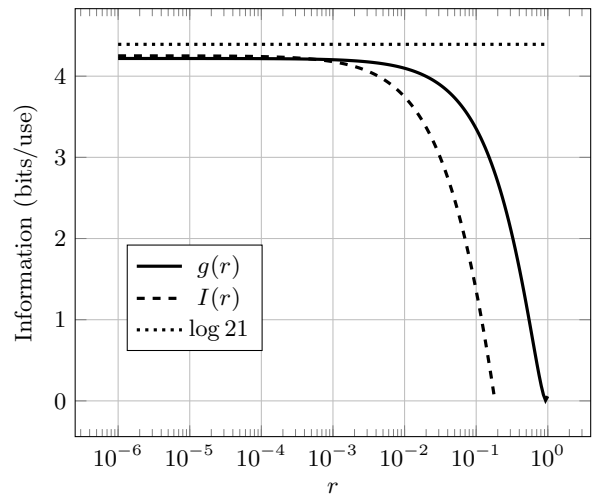

It is worth noting that each ``use" consists of a single transmission of a codon through the ribosomal channel to an amino acid.

A linear-log plot of the lower bound $g(r)$ of $\mathcal{C}$ as a function of the error probability $r$ is shown in Fig. 3. It is straightforward to show that $g(r)$ is decreasing on $(0, 20/21)$, is increasing on $(20/21,1)$, and has a root at $r=20/21$.

When $r=20/21$, Eq. (2) has the highest degree of symmetry and each amino acid is equally likely. Thus, no information is transmitted by the ribosome. Once $r$ increases beyond this point, asymmetry is reintroduced into the channel. This broken symmetry appears as a slight increase in $g(r)$ for $r \in (\frac{20}{21},1)$.

Experimentally measured error probabilities $r$ lie approximately within the range $10^{-4} \leq r \leq 10^{-3}$ \cite{edelmann77, drummond09, parker89, ogle05, kramer07, alberts22}, indicating that the capacity is bounded below by values very close to the peak of $g(r)$ (solid curve in Fig. 3). This observation is one demonstration of the ribosome's ability to translate accurately.

Since $g(r)$ is decreasing on $(0,20/21)$ and $g(0) \approx 4.2181$, we have
\begin{equation}
    4.2181 \lesssim \mathcal{C} \lesssim 4.3923 \enskip \frac{\textrm{bits}}{\textrm{use}},
    \label{eq:5}
\end{equation}

\noindent where ``$\lesssim$" indicates that $\mathcal{C}$ lies strictly between the bounds as they appear but that the bounds can be better approximated by appropriate rounding once more significant digits are taken into account.

Each codon is specified by $\log64=6$ bits, so Eq.~(\ref{eq:5}) becomes
\begin{equation}
0.7030 \lesssim \mathcal{C} \lesssim 0.7321 \enskip \frac{\textrm{codons}}{\textrm{use}}.
\label{eq:6}
\end{equation}

Evaluating $g(r)$ at $r = 1 \times 10^{-4}$, a typical value for the ribosome's error probability, yields
\begin{equation}
    0.7027 \lesssim \mathcal{C} \lesssim 0.7321 \enskip 
    \frac{\textrm{codons}}{\textrm{use}}.
    \label{eq:7}
\end{equation}

Yockey modeled the genetic communication system using a different channel matrix Eq.~(\ref{eq:2}) that incorporates point mutations \cite{yockey74, yockey05}. Using this alternative conditional probability distribution, Yockey derives the system's corresponding mutual information and through several approximations obtains $I(X;Y)= H(X)- 1.68 + 6.509r \log0.4594r$, where $H(X)$ is the entropy of input $X$. Djordjevic maximizes Yockey's expression to obtain a channel capacity using the marginal distribution of $X$ that is uniform over the non-stop codons (i.e., $p(x)=1/61$ for any non-stop codon $x$) and zero over the stop codons (i.e., $p(\text{UAA})=p(\text{UAG})=p(\text{UGA})=0$) \cite{djordjevic12}. Doing so yields $H(X)=\log 61$, so that $I(r):= I(X;Y)$ becomes
\begin{equation}
    I(r)= \log 61 - 1.68 + 6.509r \log0.4594r.
    \label{eq:8}
\end{equation}

\noindent We plot Eq.~(\ref{eq:8}) using this value (dashed curve) alongside our calculated $g(r)$ (solid curve) in Fig. 3.

As seen in Fig. 3, our result differs slightly from Eq.~(\ref{eq:8}). As stated in Theorem II.1, we predict a range of possible values for $\mathcal{C}$, with greater uncertainty as $r$ increases. It is worth noting that $g(r)$ represents a lower bound on an upper bound, namely, the capacity $\mathcal{C}$.

We recognize that, by the asymmetry of the channel (Fig. 2), we are able to analytically calculate only bounds on the capacity, such as Eqs.~(\ref{eq:5}),~(\ref{eq:6}),  and~(\ref{eq:7}). Thus, to verify the bounds in Theorem II.1 and better approximate the capacity, we do so numerically, which we outline in the next section.

\subsection{\label{sec:level2}Numerical Approximation of the Capacity}

To approximate the channel capacity, we apply the well-known Blahut-Arimoto algorithm, which is often used to compute capacities for arbitrary channels \cite{arimoto72, blahut72}.

Given an input set $\mathcal{X}$ and output set $\mathcal{Y}$, the problem of computing a channel capacity (Eq. (1)) amounts to maximizing the mutual information $I(X;Y)$ between the channel input and output over all possible input distributions $p_X$. One method for accomplishing this maximization is to calculate the gradient of $I(X;Y)$. However, this direct method often leads to a nonlinear system in a high-dimensional space. For the ribosome, this space has $|\mathcal{X}|=64$ dimensions, one dimension for each codon, leading to a computationally intractable problem.

Fortunately, the Blahut-Arimoto algorithm provides an alternative, efficient method for computing the capacity. Note that for a fixed conditional distribution $p(y|x)$, $I(X;Y)$ is a concave function of the input distribution, i.e., $I(X;Y)=I(p_X)$. The algorithm iteratively yields a sequence of input distributions $\{Q_n\}_{n \in \mathbb{N}}$. This sequence, in turn, yields a sequence of mutual informations $\{I(Q_n)\}_{n \in \mathbb{N}}$ that monotonically converges to the capacity quickly.

\begin{figure}[t]
    \centering
    \begin{tikzpicture}
\begin{axis}[
    title={},
    xlabel={Iteration $n$},
    ylabel={$I(Q_n)$ (bits/use)},
    xmin=0, xmax=12,
    ymin=4.2, ymax=4.4,
    xtick={0,1,2,3,4,5,6,7,8,9,10,11},
    ytick={4.2,4.25,4.3,4.35,4.4},
    ymajorgrids=true,
    grid=major,
]

\addplot[color=black, mark=square]
 coordinates {
    (1,4.216271304889404593572316734951275070472)(2,4.390412185189819148254745929825745426967)(3,4.390412196440942880846296091770044861574)(4,4.390412196440943377022493830311762864188)(5,4.390412196440943377022515709385097211534)(6,4.390412196440943377022515709386061977354)(7,4.390412196440943377022515709386061977392)(8,4.390412196440943377022515709386061977404)(9,4.390412196440943377022515709386061977398)(10,4.390412196440943377022515709386061977398)(11,4.390412196440943377022515709386061977405)
   };
    
\end{axis}
\end{tikzpicture}
    \caption{Results of the Blahut-Arimoto algorithm with $r=1 \times 10^{-4}$. The mutual information $I(Q_n)$ converges to the capacity $\mathcal{C}$ with monotonically.}
   \label{fig:my_label}
\end{figure}
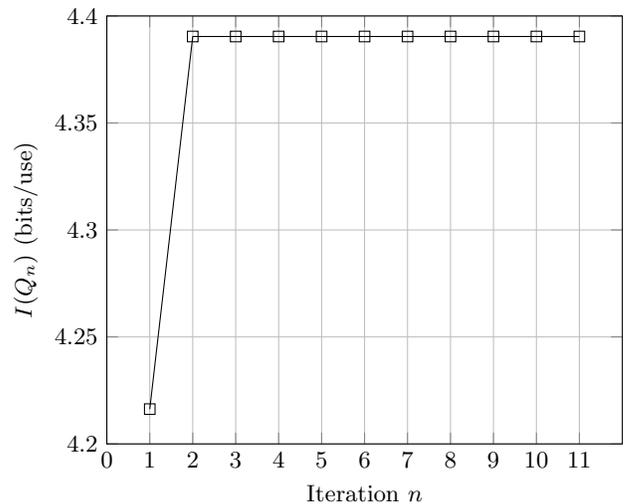

More precisely, starting from an arbitrary initial input distribution $Q_1$, for each $x \in \mathcal{X}$, the quantity
\begin{equation}
    T_{n}(x):=\sum_{y \in \mathcal{Y}}p(y|x)\log{ \bigg( \frac{Q_{n}(x)p(y|x)}{R_n(y)} \bigg)}
    \label{eq:9}
\end{equation}

\noindent is calculated for each iteration, where $R_n(y)$ is the marginal distribution of the output:
\begin{equation}
    R_n(y):= \sum_{x \in \mathcal{X}} p(y|x)Q_n(x).
    \label{eq:10}
\end{equation}

A sequence of input distributions is then calculated according to the following rule:
\begin{equation}
    Q_{n+1}(x) = \frac{e^{T_{n}(x)}}{\sum_{x' \in \mathcal{X}}e^{T_{n}(x')}}.
    \label{eq:11}
\end{equation}

It can be shown that $I(Q_n) \xrightarrow{n \to \infty} \mathcal{C}$ monotonically from below (Theorem 3 in \cite{blahut72}) and that the channel capacity $\mathcal{C}$ satisfies \cite[p. 524]{gallager68}
\begin{equation}
    m_n \leq \mathcal{C} \leq M_n,
    \label{eq:12}
\end{equation}

\noindent where
\begin{align}
    m_n&:= \min_{x \in \mathcal{X}} T_n(x)- \log Q_n(x),\\
    M_n&:= \max_{x \in \mathcal{X}} T_n(x)-\log Q_n(x).
\end{align}

Eq.~(\ref{eq:12}) provides a termination criterion that stops the algorithm once $I(Q_n)$ falls within a chosen accuracy of $\mathcal{C}$.

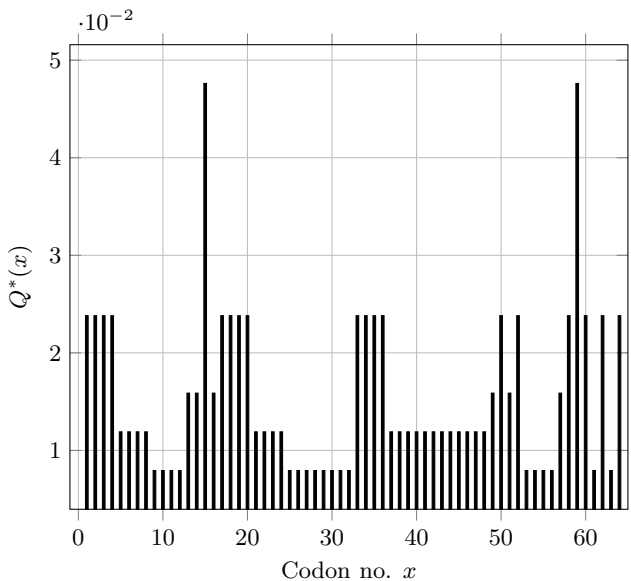
\begin{figure}[t]
\begin{tikzpicture}
 \centering
\begin{axis} [ybar,bar width=1.0pt, width= 9cm, xlabel=Codon no. $x$, xmax=65,xmin=-1,
    ylabel={$Q^*(x)$},grid]
\addplot[black,fill=black] coordinates {
    (1,0.02380952) 
    (2,0.02380952) 
    (3,0.02380952) 
    (4,0.02380952)
    (5,0.01190476)
    (6,0.01190476)
    (7,0.01190476)
    (8,0.01190476)
    (9,0.00793651)
    (10,0.00793651)
    (11,0.00793651)
    (12,0.00793651)
    (13,0.01587302)
    (14,0.01587302)
    (15,0.04761905)
    (16,0.01587302)
    (17,0.02380952)
    (18,0.02380952)
    (19,0.02380952)
    (20,0.02380952)
    (21,0.01190476)
    (22,0.01190476)
    (23,0.01190476)
    (24,0.01190476)
    (25,0.00793651)
    (26,0.00793651)
    (27,0.00793651)
    (28,0.00793651)
    (29,0.00793651)
    (30,0.00793651)
    (31,0.00793651)
    (32,0.00793651)
    (33,0.02380952)
    (34,0.02380952)
    (35,0.02380952)
    (36,0.02380952)
    (37,0.01190476)
    (38,0.01190476)
    (39,0.01190476)
    (40,0.01190476)
    (41,0.01190476)
    (42,0.01190476)
    (43,0.01190476)
    (44,0.01190476)
    (45,0.01190476)
    (46,0.01190476)
    (47,0.01190476)
    (48,0.01190476)
    (49,0.01587302)
    (50,0.02380952)
    (51,0.01587302)
    (52,0.02380952)
    (53,0.00793651)
    (54,0.00793651)
    (55,0.00793651)
    (56,0.00793651)
    (57,0.01587302)
    (58,0.02380952)
    (59,0.04761905)
    (60,0.02380952)
    (61,0.00793651)
    (62,0.02380952)
    (63,0.00793651)
    (64,0.02380952)
};
\end{axis}
\end{tikzpicture}
\caption{Capacity-achieving input distribution $Q^*$. Codons are ordered alphabetically. The codons having the largest probabilities are the start codon AUG, which codes for methionine, and UGU, which codes for cysteine.}
   \label{fig:my_label}
\end{figure}

We start by initializing $\{Q_n\}_{n \in \mathbb{N}}$ with the uniform distribution, i.e.,
\begin{equation}
    Q_1(x)= \frac{1}{64}, \enskip \forall x \in \mathcal{X}.
    \label{eq:15}
\end{equation}

We iteratively generate the subsequent terms of the sequence $\{Q_n\}_{n \in \mathbb{N}}$ using Eqs.~(\ref{eq:9}) -~(\ref{eq:11}) and the ribosome's characteristic conditional probability distribution Eq.~(\ref{eq:2}) for the error probability $r=1 \times 10^{-4}$. We allow the algorithm to continue until the difference $M_n -m_n \leq 10^{-35}$.

The results of this algorithm are plotted in Fig. 4. We observe monotonic convergence to a value $\mathcal{C}\approx4.3904$ bits/use $=0.7317$ codons/use, which falls within the bounds of Eq.~(\ref{eq:7}). In fact, these values lie very near the capacity's upper bound.

In addition to the capacity, the Blahut-Arimoto algorithm also outputs an approximation to the capacity-achieving distribution $Q^*$, that is, the input distribution for which $\mathcal{C}=I(Q^*)$. This distribution is shown in Fig. 5. It is currently unknown whether or not the approximated optimum $Q^*$ is unique.

\subsection{\label{sec:level2}The ribosome's capacity in time}

In practice, one has access to only the ribosome's \textit{in vivo} or \textit{in vitro} translation rate, which is needed so that Shannon's theorem may be applied. Time must be incorporated so that we may compare our results to experimentally measured rates.

In Ref. \cite{fluitt07}, Fluitt \textit{et al.} devise a model of translation showing that competition between cognate, near-cognate, and non-cognate aminoacyl-tRNAs (aa-tRNAs) cause delays in the observed translation rate. The authors perform Monte Carlo simulations of the ribosome during translation that yield the average translation time for each of the 64 codons \cite{fluitt07}. Their model accounts for peptide bond catalysis and the translocation of the ribosome from one codon to the next using experimentally determined kinetic rate constants obtained \textit{in vitro} for \textit{E. coli} by Gromadski and Rodnina \cite{gromadski04}.

Under the assumption that each aa-tRNA that arrives via diffusion is an aa-tRNA corresponding correctly to the codon in the ribosome’s active site  (i.e., each aa-tRNA is cognate), the authors find that the average translation time for a single codon at 37$^{\circ}$C is $\tau_{ribo}= 9.06 \enskip \textrm{ms}$. We interpret $\tau_{ribo}$ as the time that corresponds to the condition where aa-tRNAs are in sufficiently high concentrations such that aa-tRNA availability does not limit the rate of translation. That is, we take $\tau_{ribo}$ to be a good estimate of the ribosome's theoretically minimal translation time.

Dividing Eq.~(\ref{eq:7}) by $\tau_{ribo}$, we obtain the capacity range
\begin{equation}
    77.5964\lesssim \mathcal{C}^*_{ribo} \lesssim 80.7655 \enskip 
    \frac{\textrm{codons}}{\textrm{second}},
    \label{eq:16}
\end{equation}

\noindent where we have defined $\mathcal{C}^*_{ribo}:=\mathcal{C}/\tau_{ribo}$.

In addition, by dividing our numerical approximation $\mathcal{C} \approx 0.7317$ codons/use by $\tau_{ribo}$, we obtain $\mathcal{C}^*_{ribo} \approx 80.7655$ codons/s, which agrees with the range in Eq.~(\ref{eq:16}). We can also see that our approximated capacity lies very near the upper bound of the analytically calculated capacity.

Experimental ribosomal translation rates in prokaryotes fall approximately within 13-22 codons/s \cite{bremer08}. (Eukaryotic translation is even slower at about 5 codons/s \cite{hershey19, palmiter75}.) This range lies below the range of Eq.~(\ref{eq:16}) by a large margin, which implies, by Shannon's Noisy Channel Coding theorem, that the ribosome is able to translate at its observed speeds without sacrificing accuracy.

\section{\label{sec:level1}Summary \& Conclusion}

We have shown that the accuracy of the ribosome can be explained through purely information-theoretic means by introducing a new model that views the ribosome as a discrete memoryless channel. The ribosomal channel operates at rates below its capacity in time, allowing it to reliably transmit information with an arbitrary degree of error. We have shown this result by analytically bounding and numerically computing the ribosome's channel capacity and verifying that these values lie above the ribosome's experimentally observed operation rate. Our study is, as far as we know, the first to compare experimentally determined translation rates with a calculated capacity, showing that these rates lie safely below the ribosome's channel capacity.

To summarize, our result successfully explains, from an information-theoretic perspective, existing observations that the ribosome translates accurately at experimentally measured translation rates.

It is worth noting that Shannon's theorem is a nonconstructive theorem. In other words, although the theorem guarantees the existence of a coding scheme that achieves information transmission having an arbitrary degree of error, such a scheme is not specified.

It is well-known that there are many other alternative, naturally occurring genetic codes, with the standard genetic code the most prevalent \cite{osawa92}. For example, vertebral mitochondria utilize a genetic code that maps the codon AUA to the amino acid methionine, whereas the standard genetic code maps AUA to isoleucine. Our method can be extended to other genetic codes by changing the function $\mathcal{G}$ appropriately, and we anticipate this accommodation may be accomplished at a later time.

Several other questions naturally arise when considering alternative genetic codes in the context of our model. Can the channel capacity be further optimized by choosing a different genetic code? And if so, which one? Is it the standard genetic code? And as we mention above, it is currently unknown whether or not the numerically computed capacity-achieving distribution $Q^*$ is unique. These are some questions that we hope will be addressed in a future study.

The ribosome is found universally across all domains of life, albeit with some variations across these domains. Taken together, our results for the ribosome may serve as a case study of a more general feature of biological machines, namely, that biomolecules, when viewed as information channels, have evolved ways to process information quickly while minimizing errors. One such class of machines may include other enzymes such as DNA polymerases during DNA replication and RNA polymerases during transcription.

\begin{acknowledgments}
This work was partially conducted on the territories of the Kickapoo-Mascouten, Miami, Massachusett, Canarsee, Potawatomi, and Native Hawaiian (K$\overline{\mbox{a}}$naka Maoli) peoples. Many thanks to Shelley Weinberg, Brent Kirkpatrick, Pierre Albin, Tyler Earnest, Lav Varshney, Gustavo Caetano-Anoll\'{e}s, and the British Library. We are also grateful to the anonymous referees for reviewing this manuscript. K.L.K. was partially supported by National Science Foundation CAREER award DMS-1254791 and a Simons Fellowship. D.A.I. was partially supported by National Science Foundation Graduate Research Fellowship DGE-1746047.
\end{acknowledgments}

\appendix

\section{Proof of Theorem II.1}

Given random variables $X$ and $Y$ in $\mathcal{X}$ and $\mathcal{Y}$, respectively, we substitute both the standard definition of mutual information and Eq.~(\ref{eq:2}) into Eq.~(\ref{eq:1}) to obtain
\begin{align}
    \mathcal{C}&= \sup_{p_X} I(X;Y) \nonumber \\
    &= \sup_{p_X} \sum_{x \in \mathcal{X}} \sum_{y \in \mathcal{Y}} p(x,y)\log \frac{p(x,y)}{p_X(x)p_Y(y)} \nonumber \\
                &= \sup_{p_X} \sum_{x \in \mathcal{X}}\sum_{y\in \mathcal{Y}} p(y|x)p(x)\log \frac{p(y|x)}{p(y)},
                \label{eq:A1}
\end{align}

\noindent where $p(x):=p_X(x)$ and $p(y):=p_Y(y)$ are the respective marginal distributions for $X$ and $Y$, and $p(x,y)$ is their corresponding joint distribution.

The estimate continues by picking a particular marginal probability distribution $p_X$, namely the uniform distribution over the 64 possible codons, which gives a lower bound on the supremum as follows:
\begin{align}
    \mathcal{C} &= \sup_{p_X} \sum_{x \in \mathcal{X}} \sum_{y \in \mathcal{Y}} p(y|x)p(x)\log\bigg( \frac{p(y|x)}{\sum_{x'}p(y|x')p(x')}\bigg) \nonumber \\
                &\geq \frac{1}{64} \sum_{x \in \mathcal{X}} \sum_{y \in \mathcal{Y}} p(y|x)\log\bigg( \frac{p(y|x)}{\frac{1}{64}\sum_{x'}p(y|x')}\bigg) \nonumber \\
                &=: g(r).
                \label{eq:A2}
\end{align}

To simplify notation, we define $f(y)$ by
\begin{equation}
    f(y) := \sum_{x \in \mathcal{X}} p(y|x) \log \bigg( \frac{64 p(y|x)}{\sum_{x'}p(y|x')} \bigg),
    \label{eq:A3}
\end{equation}

\noindent so that
\begin{equation}
    g(r)=\frac{1}{64} \sum_{y \in \mathcal{Y}} f(y).
    \label{eq:A4}
\end{equation}

$\mathcal{G}$ is not injective (i.e., the genetic code is degenerate), and amino acids can be grouped according to the number of codons that map to each amino acid. (A table of the standard genetic code can be found in many standard textbooks in biology, such as \cite{clark19}.) For example, two amino acids---methionine (Met) and tryptophan (Trp)---both have exactly one codon that map to each, whereas three other amino acids---leucine (Leu), serine (Ser), and arginine (Arg)---have exactly six such codons each. Therefore, $f(\text{Met})=f(\text{Trp})$, and so on. Using Eq.~(\ref{eq:2}), for Met we have
\begin{equation}
    \sum_{x' \in \mathcal{X}} p(\text{Met}| x') = (1-r) + \frac{63r}{20}= \frac{43r+20}{20}.
    \label{eq:A5}
\end{equation}

\noindent Here, the term $1-r$ corresponds to the case $y=\mathcal{G}(x)$, and the term $\frac{63r}{20}$ corresponds to the cases where $y \neq \mathcal{G}(x)$. Substituting Eq.~(\ref{eq:A5}) into Eq.~(\ref{eq:A3}), we have
\begin{align}
    f(\text{Met})&= \sum_{x \in \mathcal{X}} p(\text{Met}|x) \log \bigg( \frac{1280 p(\text{Met}|x)}{43p+20} \bigg) \nonumber \\
    &= q \log \frac{1280q}{43r+20} + \frac{63r}{20}\log \frac{64r}{43r+20},
    \label{eq:A6}
\end{align}

\noindent where $q:=1-r$.

$f(y)$ for each of the other amino acids is calculated similarly. Doing so for each amino acid $y \in \mathcal{Y}$ and substituting the results into Eq.~(\ref{eq:A4}), we obtain
\begin{align}
    g(r) &= \frac{1}{64} \bigg\{2q\log \frac{1280q}{43r+20}+\frac{63r}{10} \log \frac{64r}{43r+20} \nonumber \\
         & + 18q\log \frac{640q}{11r+20}+\frac{279r}{10} \log \frac{32r}{11r+20} \nonumber \\
         & + 6q\log \frac{1280q}{r+60}+\frac{61r}{10} \log \frac{64r}{r+60}\nonumber \\
         & + 20q\log \frac{64q}{4-r}+15p\log \frac{16r}{5(4-r)}\nonumber \\
         & + 18q\log \frac{640q}{60-31r}+\frac{87r}{10} \log \frac{32r}{60-31r} \bigg\}. \label{eq:A7}
\end{align}

Combining Eqs.~(\ref{eq:A2}), ~(\ref{eq:A7}), and~(\ref{eq:3}), we obtain the desired result.

\bibliography{main}

\end{document}